\documentclass[11pt]{article}
\usepackage{xspace,amssymb,amsmath,helvet,graphicx}
\begin{document}
\newcommand{\dee}{\,\mbox{d}}
\newcommand{\naive}{na\"{\i}ve }
\newcommand{\eg}{e.g.\xspace}
\newcommand{\ie}{i.e.\xspace}
\newcommand{\pdf}{pdf.\xspace}
\newcommand{\etc}{etc.\@\xspace}
\newcommand{\PhD}{Ph.D.\xspace}
\newcommand{\MSc}{M.Sc.\xspace}
\newcommand{\BA}{B.A.\xspace}
\newcommand{\MA}{M.A.\xspace}
\newcommand{\role}{r\^{o}le}
\newcommand{\signoff}{\hspace*{\fill} Rose Baker \today}
\newenvironment{entry}[1]%
{\begin{list}{}{\renewcommand{\makelabel}[1]{\textsf{##1:}\hfil}%
\settowidth{\labelwidth}{\textsf{#1:}}%
\setlength{\leftmargin}{\labelwidth}
\addtolength{\leftmargin}{\labelsep}
\setlength{\itemindent}{0pt}
}}%
{\end{list}}
\title{A new asymmetric generalisation of the t-distribution}
\author{R. D. Baker\\School of Business\\University of Salford, UK\\email rose.baker@cantab.net}
\maketitle
\begin{abstract}
A 6-parameter fat-tailed distribution is proposed that generalises the t-distribution and allows asymmetry of scale and also of tail power, whilst avoiding the discontinuity of
the second derivative of the split-t (AST) distribution. With the sixth parameter set to unity and no asymmetry, the distribution reduces to a t-distribution,
but with the sixth parameter reduced, fatter tails than those of the t-distribution are allowed (the tails start earlier) and the distribution generalises Johnson's $S_U$ distribution. 
Data fitting is illustrated with examples.
\end{abstract}
\section*{Keywords}
skew-t distribution; arcsinh transformation; AST distribution.

\section{Introduction}
There is a widespread problem of modelling and doing inference when the distribution of the variable of interest has long tails and is skewed. This occurs commonly in finance,
\eg for returns, but also in the life sciences, telecommunications, etc. The t-distribution has polynomial tails, and has been widely used in modelling fat-tailed data. 
The problem then is to generalise the t-distribution so that it can exhibit skewness. 
After a brief account of previous attempts to do this, a new distribution that does so is described in the next section.

The asymmetric fat-tailed distribution of returns etc in finance has been modelled by the generalised asymmetric t distribution (AST)(\eg Zhu and Galbraith, 2010, 2011)
The pdf is $f(x) \propto\{1+((x-\mu)/c\phi)^2/(\nu/r)\}^{-(\nu/r+1)/2}$
for $X < \mu$ and $f(x) \propto \{1+(c(x-\mu)/\phi)^2/(\nu r)\}^{-(\nu r+1)/2}$
for $X > \mu$ (our notation). This form gives continuity of $f(x)$ and its
first derivative (zero) at $X=\mu$.
Here for the random variable $X > \mu$ the pdf is proportional to the pdf of a t-distribution centred at $\mu$ with scale $c\phi$ and degrees of freedom $\nu r$,
while for $X < \mu$ the pdf is proportional to that of a t-distribution with scale $\phi/c$, degrees of freedom $\nu/r$. The two constants of proportionality are chosen so that the total probability is unity,
and there is no discontinuity at the join. Data can exhibit two types of skewness:
asymmetry in scales so that $c \ne 1$ is usually seen, and sometimes asymmetry in tail behaviours
$f(x) \propto x^{-(\nu r+1)}$ as $x \rightarrow \infty$ and $f(x) \propto |x|^{-(\nu/r+1)}$ as $x \rightarrow -\infty$ is also found, so that $r \ne 1$.
This distribution generalises the earlier skew-t distribution of Hansen (1994) and Fernandez and Steel (1998) where the two tail powers are the same.

Other  skew-t distributions that include the t-distribution as a special case are those of Branco and Dey (2001) and Azzalini and Capitanio (2003). This distribution has the same tail behaviour in both tails,
as does that of Jones and Faddy (2003). Rosco, Jones and Pewsey (2011) apply their sinh-arcsinh transformation (Jones and Pewsey, 2009) to the t-distribution 
so that the transformed t-variable is skew. However, the only distribution that allows both types of skewness is the AST distribution.

The AST distribution has the drawback of the gluing together of two disparate halves, which is unaesthetic. It also seems unlikely that such a construct could reflect actual behaviour of the data,
and the discontinuity in the second derivative of the log-likelihood function means that the usual regularity
conditions for maximum likelihood estimation are not satisfied, and makes inference for parameter values difficult. Experience reveals no problem
in fitting the distribution by likelihood maximisation, but the estimation of standard errors on fitted model parameters is problematical because it relies on the second derivative of the log-likelihood.

We therefore sought a more natural fat-tailed distribution that possessed both types of asymmetry, and were fortunate to find one that generalises the t-distribution.
\section{A new asymmetric distribution}
In motivating the new distribution, it is perhaps easiest to start with the  type IV generalised logistic distribution (Johnson, Kotz and Balakrishnan, 1995, p142). This has pdf
$f(y) \propto \frac{\exp(-qy)}{(1+\exp(-y))^{p+q}}$, and it is also the log-F distribution (the logarithm of a random variate from the F distribution). This distribution
allows asymmetry, the right tail behaving as $f(y) \propto \exp(-qy)$ and the left as $f(x) \propto \exp(-p|y|)$. Tadikamalla and Johnson (1982) and Johnson and Tadikamalla (1992) applied Johnson's arcsinh transformation
to the logistic distribution $f(y)=\exp(-y)/(1+\exp(-y))^2$, so converting the exponential tail into a power-law tail and obtaining a fat-tailed distribution, but we apply the arcsinh transformation to the type IV logistic distribution,
obtaining a 6-parameter distribution with the required properties.

We start then with the type IV rescaled generalised logistic pdf, in our notation
\begin{equation}f(y)=\frac{\alpha (1+r^2)}{r}\frac{\{\exp(\alpha r y)+\exp(-(\alpha/r)y)\}^{-\nu/\alpha}}{B(\frac{\nu/\alpha}{1+r^2},\frac{r^2\nu/\alpha}{1+r^2})},\label{eq:y}\end{equation}
where $B$ denotes the beta function. The parameter $r$ controls the asymmetry, with $r=1$ for a symmetric distribution. Now we apply the arcsinh transformation,
where 
\begin{equation}y=\ln c+\sinh^{-1}((x-\mu)/\phi)\equiv \ln c+\ln\{(x-\mu)/\phi+\sqrt{1+((x-\mu)/\phi)^2}\}.\label{eq:trans}\end{equation}
To obtain the pdf of $X$ we use the relation between pdfs $f_x(x)=f_y(y)|\dee y/\dee x|$, where the Jacobian $\dee y/\dee x=(1/\phi)/\sqrt{1+((x-\mu)/\phi)^2}$,
Writing for brevity 
\[cg((x-\mu)/\phi)=(x-\mu)/\phi+\sqrt{1+((x-\mu)/\phi)^2},\] the pdf is
\begin{equation}f(x)=\frac{\alpha (1+r^2)}{r\phi}\frac{\{(cg((x-\mu)/\phi))^{\alpha r}+(c g((x-\mu)/\phi))^{-\alpha/r}\}^{-\nu/\alpha}}{B(\frac{\nu/\alpha}{1+r^2},\frac{r^2\nu/\alpha}{1+r^2})}(1+((x-\mu)/\phi)^2)^{-1/2}.\label{eq:x}\end{equation}
Here $\nu > 0$ controls tail power, $\mu$ is in a sense the centre of location (but not necessarily the mean, which may not exist), $\phi > 0$ is a measure of scale (but not the variance, which may not exist),
$r > 0$ controls tail power asymmetry, $c > 0$ controls the scale asymmetry, and $\alpha > 0$ controls how early `tail behaviour' is apparent. We propose calling this new distribution the GAT distribution (generalised asymmetric t).
A gat is a gun in obsolete American slang, so the idea here is that this distribution, with its 6 parameters, is a powerful weapon.

To explore the meaning of these parameters, let $x$ become large, when $g(x) \approx 2cx/\phi$, so that $f(x) \approx \frac{\alpha(1+r^2)}{r\phi B}(2cx/\phi)^{-r\nu}(x/\phi)^{-1}$, \ie $f(x) \propto x^{-\nu r-1}$.
Since $g((x-\mu)/\phi))g(-(x-\mu)/\phi))=1$, when $x$ is large and negative $f(x) \approx \frac{\alpha(1+r^2)}{r\phi B}(2|x|/c\phi)^{-\nu/r}(|x|/\phi)^{-1}$, \ie $f(x) \propto |x|^{-\nu/r-1}$.
This shows how $\nu$, $r$ and $c$ control tail behaviour; $c$ controls the ratio of probability masses in each tail and $r$ controls the ratio of powers of $x$ in each tail.

Setting the asymmetry parameters $r=1, c=1$ and setting $\alpha=1$ the pdf becomes
\[f(x)=(2/\phi)\{(1+((x-\mu)/\phi)^2\}^{-{(\nu+1)/2}}/B(\nu/2,\nu/2),\]
so that on applying the Legendre duplication formula to regain a more familiar form of the constant, we see that $X$ is a rescaled random variate following the t distribution with $\nu$ degrees of freedom; specialising to $\phi=\sqrt{\nu}$ and $\mu=0$ we obtain the standard t distribution.
This is a restatement of the fact noted by several authors (Johnson, Kotz and Balakrishnan, p346) that for an $F$ distribution with $\nu$ and $\nu$ degrees of freedom, $\frac{\sqrt{\nu}}{2}(F^{1/2}_{\nu,\nu}-F^{-1/2}_{\nu,\nu})$ follows a t distribution with $\nu$ degrees of freedom.

With $\alpha=1$ we have a 5-parameter distribution that turns out to fit returns data almost identically well to the AST distribution, but which does not have the
same inferential problems, as the log-likelihood function has no discontinuities in derivatives. The GAT distribution can fit very skew data, and the only area where it
cannot reproduce the behaviour of the AST distribution is the case where $\nu\rightarrow\infty$. The GAT distribution cannot be skew when both tails are normally distributed,
but the AST distribution can.

On allowing the sixth parameter $\alpha$ to vary from unity,
we have a more general distribution that sometimes fits the data better. As $\alpha$ increases, the fatness of the tails decreases, while the power-law behaviour remains the same.
Thus this distribution allows us to fit financial and other data with fatter tails even than the t-distribution can cope with, by floating $\alpha$ below unity.

As $\alpha \rightarrow 0$ and $\nu \rightarrow\infty$ such that $\eta=\nu\alpha$ remains constant, by expanding the pdf about the mode it is readily seen that
$Y$ is normally distributed, with mean zero and variance $\eta^{-1}$. Hence $X$ follows Johnson's $S_U$ distribution. In this case the distribution of $X$ has the tail behaviour of a lognormal distribution.

It is possible to ignore the Jacobian in (\ref{eq:x}), and so obtain a slightly simpler distribution, for which the mode is easily found as a transformation of the mode of $Y$,
which is $y_0=\frac{-2\ln(r)}{\alpha(r+1/r)}$.
The mode of the GAT distribution on the other hand must be found iteratively, \eg by using Newton-Raphson iteration from the corresponding value of $x$,
\[x_0=\mu+(\phi/2)\{r^{\frac{-2}{\alpha(r+1/r)}}/c-cr^{\frac{2}{\alpha(r+1/r)}}\}.\]
This distribution fitted data equally well. However, without the Jacobian, 
the minimum value of $\nu$ for which the pdf can be defined is $\nu > \text{max}(r,1/r)/\alpha-1$, which is messy. Hence we do not develop this distribution further.
It does however have broadly similar properties to the GAT distribution, and could also be useful.

\section{Moments, distribution function and random numbers}
The moments and distribution function can be expressed analytically in terms of beta functions.
We shall need the (complete) beta function $B(a,b)=\int_0^1 q^{a-1}(1-q)^{b-1}\dee q$,
the incomplete beta function $B_I(a,b;c)=\int_0^c q^{a-1}(1-q)^{b-1}\dee q$ and the regularised incomplete beta function
$B_R(a,b;c)=B_I(a,b;c)/B(a,b)$.

 It is convenient to start by showing that the pdf integrates to unity. This and the moments are best computed by
working with $f_y(y)$ rather than $f_x(x)$, and then writing $x=\mu+\phi\sinh(y-\ln(c))$. In (\ref{eq:y})  we can change variable of integration first to $z=y-\ln(c)$
so that now $x=\phi(\exp(z)- \exp(-z))/2$, then to $q(z)=\{1+\exp(-\alpha(r+1/r)z\}^{-1}$. Clearly $0 < q < 1$, and we have that
$z=-\frac{\ln((1-q)/q)}{\alpha(r+1/r)}$, and the Jacobian $\dee z/\dee q=q^{-1}(1-q)^{-1}/(\alpha(r+1/r))$. The normalisation given in (\ref{eq:x}) follows,
and the mean is
\begin{equation}\text{E}(X)=\mu+\phi\{
\frac{c^{-1}B(\frac{\nu}{\alpha(1+r^2)}+\delta,\frac{\nu r^2}{\alpha(1+r^2)}-\delta)-cB(\frac{\nu}{\alpha(1+r^2)}-\delta,\frac{\nu r^2}{\alpha(1+r^2)}+\delta)}{
2B(\frac{\nu}{\alpha(1+r^2)},\frac{\nu r^2}{\alpha(1+r^2)})}\}\label{eq:mean}\end{equation}
when $\nu > \max(r,1/r)$ and where $\delta=r/(\alpha(1+r^2))$. Higher moments may be computed
similarly, the $n$th moment existing when $\nu > n$ for the
symmetric distribution, else $\nu > n\max(r,1/r)$.
The algebra rapidly becomes tedious but not difficult, for example the variance can be derived as $\text{E}\{(X-\mu)^2\}-\text{E}\{(X-\mu)\}^2$, where
\begin{equation}\text{E}\{(X-\mu)^2\}=
\phi^2[c^{-2}B(\frac{\frac{\nu}{\alpha(1+r^2)}+2\delta,\frac{\nu r^2}{\alpha(1+r^2)}-2\delta)+c^2B(\frac{\nu}{\alpha(1+r^2)}-2\delta,\frac{\nu r^2}{\alpha(1+r^2)}+2\delta)}{
4B(\frac{\nu}{\alpha(1+r^2)},\frac{\nu r^2}{\alpha(1+r^2)})}]-\phi^2/2.\label{eq:variance}\end{equation}
In general, 
\[\text{E}(X-\mu)^n=(\phi/2)^n\frac{\sum_{m=0}^n (-1)^m {n \choose m}c^{n-2m}B\{\frac{\nu}{\alpha(1+r^2)}-(n-2m)\delta,\frac{\nu r^2}{\alpha(1+r^2)}+(n-2m)\delta\}}
{B(\frac{\nu}{\alpha(1+r^2)},\frac{\nu r^2}{\alpha(1+r^2)})}.\]

Using the same algebra, the distribution function $F(x)$ is the
regularised incomplete beta function
\[F(x)=B_R(\frac{\nu}{\alpha(1+r^2)},\frac{\nu r^2}{\alpha(1+r^2)}; q(x))\] 
where 
\[q(x)=\frac{1}{1+c^{-\alpha(1+r^2)/r}\{\frac{(x-\mu)}{\phi}+\sqrt{1+\frac{(x-\mu)^2}{\phi^2}}\}^{-\alpha(1+r^2)/r}}.\]
Quantiles may be found by Newton-Raphson iteration starting from $x=\mu$.

Random numbers may best be generated from the GAT distribution
by generating them from the corresponding Beta distribution, and
then transforming so that 
\[X=\mu+(\phi/2)\{c^{-1}(q/(1-q))^{\delta}-c(q/(1-q))^{-\delta}\}.\]

Distribution functions and random variates are useful for a variety of purposes, \eg the distribution function is needed for EDF-based goodness of fit tests such as the Anderson-Darling test,
and random numbers are needed for MCMC, and also in finding the distribution and p-value of a goodness of fit statistic via the parametric bootstrap.

We also have the mean absolute deviation
\[\text{E}|X-\text{E}(X)|=\mu(1-2F(\text{E}(X)))+\text{E}(X-\mu)-\]
\[\phi\{\frac{c^{-1}B_I(\frac{\nu}{\alpha(1+r^2)}+\delta,\frac{\nu r^2}{\alpha(1+r^2)}-\delta; q(\text{E}(X)))-cB_I(\frac{\nu}{\alpha(1+r^2)}-\delta,\frac{\nu r^2}{\alpha(1+r^2)}+\delta; q(\text{E}(X)))}{B(\frac{\nu}{\alpha(1+r^2)},\frac{\nu r^2}{\alpha(1+r^2)})}\}\]

In finance, the Value-at-Risk, VaR, is a quantile such that given a distribution of gains, the probability of exceeding a loss, $-X$ is $\gamma$, where \eg $\gamma=0.02$.
Then $F(-\text{VaR})=\gamma$, and VaR may be found like any quantile, by Newton-Raphson iteration using the distribution function $F$.
The expected shortfall (ES) is the expected loss given that the loss is at least the VaR, so that $\text{ES}=-\text{E}(X|X < -\text{VaR})$.
We have
\[\text{ES}=-\mu-\phi\{\frac{c^{-1}B_I(\frac{\nu}{\alpha(1+r^2)}+\delta,\frac{\nu r^2}{\alpha(1+r^2)}-\delta;q(-\text{VaR}))-cB_I(\frac{\nu}{\alpha(1+r^2)}-\delta,\frac{\nu r^2}{\alpha(1+r^2)}+\delta;q(-\text{VaR}))}{2B_I(\frac{\nu}{\alpha(1+r^2)},\frac{\nu r^2}{\alpha(1+r^2)};q(-\text{VaR}))}\}\]
so that given the VaR, ES can be computed using the incomplete beta function.

\section{Inference}
There are several points to consider regarding statistical use of the new distribution.
First, any problems arising with estimating standard errors on model parameters with the AST distribution do not arise here, as the GAT distribution has all derivatives continuous. 
Next, there is a well-known problem with Azzalini's skew-normal distribution (\eg, Hallin and Ley, 2014). This is that the derivative of the log-likelihood with respect to the skewness parameter
is zero when the parameter is zero (the skew-normal reduces to a normal distribution). This problem does not occur here.

Another is the sequence of model fitting; with so many parameters, how do we proceed? Experience shows that many skew distributions require only the parameters $\mu, \phi, \nu$ and $c$, so that skewness is modelled purely
by having different probability mass in the two tails. We can set $r=\alpha=1$. For some financial data the tail powers are different, so we can then try $r \ne 1$.
It is also worth then trying $\alpha \ne 1$ as this sometimes improves the fit. The 4-parameter distribution with parameters $\eta=\nu\alpha$, $\mu, \phi$ and $c$
is sometimes a good fit, where $\nu\rightarrow\infty$; this is Johnson's $S_U$ distribution. In practice, to fit this 4-parameter distribution given the ability to fit the full GAT distribution, one can set for example $\nu=200$
and vary $\mu, \phi,\alpha$ and $c$. It is common in frequentist inference to choose the best model, for example the minimum-AIC model.
A Bayesian approach would instead put prior distributions on all the parameters.

Another inferential issue is that of carrying out regressions on covariates. With a skew distribution it is often not clear which measure of central tendency should be modelled as a function of covariates, \eg the mean, median, or mode.
Here it is suggested that the parameter $\mu$ be modelled. From (\ref{eq:mean}) the mean is $\mu$ with an added function of the other five parameters.
Thus the regression for $\mu$ is also a regression for the mean, once the term for $\text{E}(X)-\mu$ from (\ref{eq:mean}) has been added to constant term of the regression.
Similarly it is also a  regression for the median and mode.
From (\ref{eq:variance}) the higher moments do not depend on $\mu$, and the variance is proportional to $\phi^2$. Thus if modelling the variance as well as the mean,
$\mu$ and $\phi$ can be modelled. These attractive results follow because the GAT distribution is a location-scale distribution.

R codes to compute the pdf, distribution function, quantiles etc are available by email from the author.

\section{Examples}
\begin{figure}
\centering
\makebox{\includegraphics{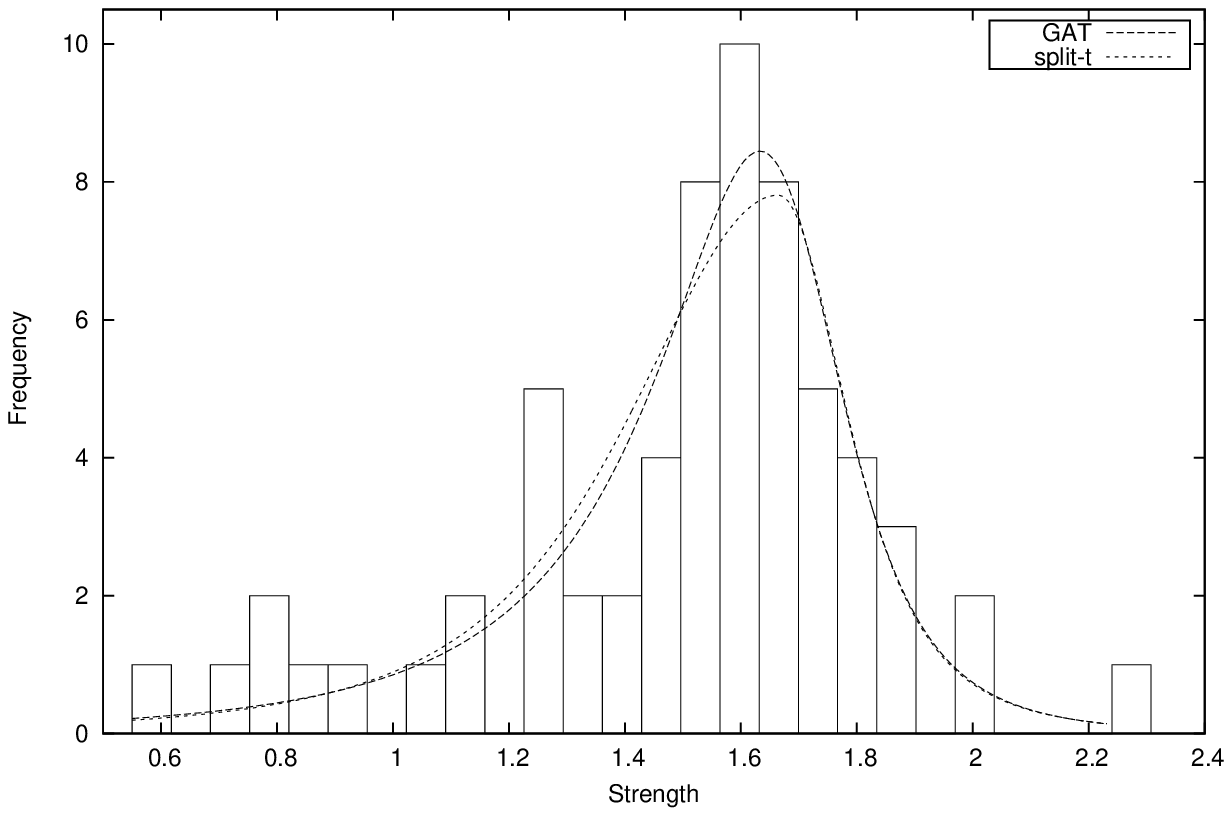}}
\caption{\label{glass} Glass strength data with fitted GAT distribution and split-t (AST) distribution.}
\end{figure}

Figure \ref{glass} shows the GAT distribution and the AST distribution fitted to data on the breaking strengths of 63 glass fibres. This dataset, first published by Smith and Naylor (1987),
has been widely used by developers of skew distributions, such as Jones and Faddy (2003), Azzalini and Capitanio (2003), Ma and Genton (2004) and Jones and Pewsey (2009). Because it is skew to the left,
it cannot be fitted well by lifetime distributions. Fitting the GAT distribution with $\alpha=r=1$ (4 parameters floated) gave minus log-likelihood $-\ell=11.7557$, compared with $11.7921$ for the AST distribution.
This shows a (very) slightly better fit, which is typical but not inevitable. Allowing tail power ($r$) to float hardly improved the fit for either distribution. However, on allowing $\alpha$ to vary from unity, 
$\alpha$ became tiny and $\nu$ large, and the fit improved slightly with $\ell=-11.207$ (shown in the figure). Hence Johnson's $S_U$ distribution, a special case of the GAT distribution, gives the best 4-parameter fit.

\begin{figure}
\centering
\makebox{\includegraphics{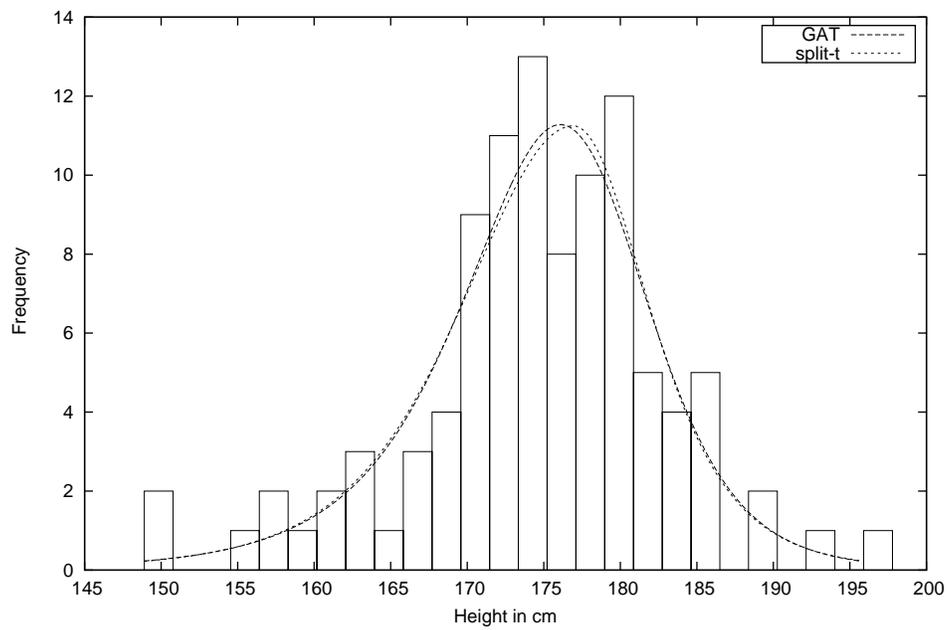}}
\caption{\label{athlete} Height of female athlete data with fitted GAT distribution and split-t (AST) distribution.}
\end{figure}

Figure \ref{athlete} shows fits to the heights in centimetres of 100 female athletes. The data are from Cook and Weisberg (1994) and have been used by Arellano-Valle {\em et al} (2004) 
to illustrate the Azzalini skew-normal distribution. Again, the plot is skew to the left. Here, the split-t and GAT distributions with $\alpha=r=1$ perform very similarly,
with $\ell=-348.3739$ for the GAT and $-348.4578$ for the split-t. Again the GAT distribution fits (very slightly) better. Here, floating $r$ and $\alpha$ does not improve the
fit, although the best 4-parameter distribution is again Johnson's $S_U$ ($\ell=-348.2031$).

Tables \ref{tab1} and \ref{tab2} show the results of maximum-likelihood fits to daily FTSE-100 and Nikkei 225 returns, respectively.
Again, in both cases the 4-parameter GAT distribution does slightly better than the split-t distribution. In the case of the Nikkei, Johnson's $S_U$ distribution
gives a significantly better fit than the split-t distribution.

There are financial data where tail powers differ, and the parameter $r$ is needed, but these simple examples really just demonstrate two points.
One is that the GAT distribution can fit at least as well as the split-t distribution. The other is that allowing the parameter $\alpha$ to vary sometimes improves the fit,
an option not available with the split-t distribution.
\section{Final comments}
The GAT distribution generalises the t-distribution in three ways: through the two types of skewness (parameters $c$ and $r$) and through how soon `tail behaviour' starts
(parameter $\alpha$). Its use in inference has been illustrated and described. It is natural to ask whether the distribution can be made multivariate. To date
no attractive way of doing this has presented itself. Many multivariate distributions, such as the multivariate t distribution, have the drawback that when the correlation 
between variables is zero, they are still not independent. Use of a copula allows multivariate distributions of variables that can be truly independent as well as uncorrelated.
Currently this is the only way to generate a multivariate-GAT distribution.
\section*{Acknowledgements}
I would like to thank Prof. Ian McHale, Dr Dan Jackson, and Prof. Chris Jones,  whose comments have improved this paper.

\section*{Figures and tables}
\begin{table}[h]
\begin{tabular}{|l|c|c|c|c|c|} \hline
Dist.  &$\nu$ & $c$ & $r$ & $\alpha$ & $-\ell$\\ \hline
GAT & 3.583&1.109& 1 (f) & 1 (f) & 11274.835 \\ \hline
AST & 3.587&1.054 & 1 (f) &-&11275.810\\ \hline
GAT & 3.572 & 1.175& 0.974 & 1 (f) & 11274.795 \\ \hline
AST & 3.593&1.035&1.050&-&11275.419\\ \hline
GAT & 3.875&1.101& 1 (f) & 0.7315 & 11274.736\\ \hline
GAT & 3.810& 1.150 & 0.981 & 0.743 & 11274.715 \\ \hline
\end{tabular}
\caption{\label{tab1}Minus the logarithm of the likelihood function for fits to the FTSE-100 returns (8028 values) and fitted parameter values. Parameter
values with (f) following were fixed at that value.}
\end{table}
\begin{table}[h]
\begin{tabular}{|l|c|c|c|c|c|} \hline
Dist.  &$\nu$ & $c$ & $r$ & $\alpha$ & $-\ell$\\ \hline
GAT&3.559&1.0886& 1(f) & 1 (f) & 12933.592 \\ \hline
AST & 3.561&1.046& 1 (f) & -&12933.857 \\ \hline
GAT&3.554&1.138&0.980 & 1 (f) & 12933.562 \\ \hline
AST & 3.564 & 1.038&1.021&-&12933.788 \\ \hline
GAT& 200 (f) & 1.073& 1 (f) & 0.00806 & 12926.376 \\ \hline
GAT& 161.7 & 1.043&1.0003&0.00996 & 12926.376 \\ \hline
\end{tabular}
\caption{\label{tab2}Minus the logarithm of the likelihood function for fits to the Nikkei 225 returns (7581 values) and fitted parameter values. Parameter
values with (f) following were fixed at that value.}
\end{table}
\end{document}